\documentclass[prl,twocolumn,aps,showpacs]{revtex4}
\usepackage{graphicx}
\begin{document}
\title{Phase Diagram of Na$_x$CoO$_2$ Studied by Gutzwiller Density
Functional Theory}

\author{Guang-Tao Wang$^1$, Xi Dai$^1$, Zhong Fang$^{1}$}

\affiliation{$^1$Beijing National Laboratory for Condensed Matter
Physics, and Institute of Physics, Chinese Academy of Sciences,
Beijing 100080, China}

\date{\today}   
\begin{abstract}
The ground states of Na$_x$CoO$_2$ ($0.0<x<1.0$) is studied by the
LDA+Gutzwiller approach, where charge transfer and orbital
fluctuations are all self-consistently treated {\it ab-initio}.  In
contrast to previous studies, which are parameter-dependent, we
characterized the phase diagram as: (1) Stoner magnetic metal for
$x>0.6$ due to $a_{1g}$ van-Hove singularity near band top; (2)
correlated non-magnetic metal without $e_g^{\prime}$ pockets for
$0.3<x<0.6$; (3) $e_g^{\prime}$ pockets appear for $x<0.3$, and
additional magnetic instability involves. Experimental quasi-particle
properties is well explained, and the $a_{1g}$-$e_g^{\prime}$
anti-crossing is attributed to spin-orbital coupling.
\end{abstract}

\pacs{71.20.-b, 71.27.+a} \maketitle

Transition-metal oxides have complex phase diagrams due to the
interplay between the charge, spin and orbital degrees of freedom.
Among them Na$_{x}$CoO$_{2}$ is a typical system showing
doping-dependent phase control~\cite{NaCoO_phase}. It has been
found experimentally as non-magnetic (NM) metal for Na poor side,
while Curie-Weiss metal for Na rich side~\cite{NaCoO_phase,Spin},
and A-type (layered) anti-ferromagnetic (AF) state for $x\sim
0.75$~\cite{A-AF}. In addition, superconductivity is discovered
for hydrated Na$_{x}$CoO$_{2}$$ \cdot y$H$_{2}$O ($x\sim
0.35$)~\cite{SC}, and charge-spin-orbital ordered states are
suggested for $x=0.5$ due to Na ordering~\cite{x=0.5}. The rich
properties of Na$_{x}$CoO$_{2}$ attract much of the
research interests due to not only its potential applications, but
also the challenging theoretical issues in this system generated
by both the \textit{multi-orbital} nature and strong e-e
correlation.

Na$_{x}$CoO$_{2}$ is crystallized in planar triangle lattice, with
each Co site being coordinated by edge-shared oxygen-octahedron.
The $e_{g}$ states are about 2eV higher than $t_{2g}$, and the
Fermi level is located within the Co-$t_{2g}$ multiplet, which
splits again into one $a_{1g}$ and two $e_{g}^{\prime }$ orbitals
under trigonal crystal field. For Na concentration $x$, the
effective number of $t_{2g}$ electrons per Co is given as 5+$x$,
and thus the low energy physics here is dominated by the multiple
orbits ($a_{1g} $+$e_{g}^{\prime }$), where charge, spin and
orbital degrees of freedom are all active. The rigorous
computational tools for such systems are still lacking, and the
observed rich phenomena remains far from even qualitative being
understood. The main controversial issues are: (1) for $x=0.3$,
are there any Fermi surface pockets for the $e_{g}^{\prime }$
band? These pockets are predicted by LDA (local density
approximation) calculations~\cite{Singh} but not observed by
ARPES~\cite{ARPES}. (2) Is the $x>0.5$ side more \textquotedblleft
correlated\textquotedblright\ than $x<0.5$ side, as suggested by
the Curie-Weiss behavior for the Na rich side? It is expected from
simple band picture that $x=1.0$ end compound is a band insulator
rather than Mott insulator.

Both LDA and LDA+$U$ methods fail for such system due to the
insufficient treatment of electron correlation. This is why many
issues look controversial following the LDA pictures. For
instances, LDA predicts ferromagnetic metal as the ground state
for the whole doping region, and LDA+$U$ even enhances the
tendency to be ferromagnetic~\cite{Louie}; the band width obtained
by LDA or LDA+$U$ is about two times larger than what observed by
ARPES~\cite{ARPES2}; the $e_g^\prime$ pockets problem as mentioned
above; and etc. To treat the electron correlation more precisely,
the Gutzwiller~\cite{ZQWang} and DMFT (dynamic mean field
theory)~\cite{Ishida} approaches has been adopted, where
fluctuation effects are included. However, those studies are only
focused on the $x=0.3$ compound using tight-binding Hamiltonian
extracted from LDA, and conflicting results are drawn due to
different parameters~\cite{DMFT-new}.

In this paper, we show that the above mentioned theoretical
challenging issues of this multi-orbital correlated electron
systems can be well studied by using the recently developed
LDA+Gutzwiller method~\cite{LDA+G}, which keeps the parameter-free
character of density functional theory (DFT) and includes all
possible charge transfer and orbital fluctuation effects
self-consistently. As the results, a phase diagram for the whole
doping region is constructed, which removes away most of the above
mentioned controversial issues, and suggests that physics here
dominates by the doping-dependent orbital, charge and spin
fluctuations.

To overcome the problem of LDA, the common procedure of LDA+$U$ and
DMFT schemes~\cite{DMFT} are to draw out from the LDA Hamiltonian the
interaction terms for the localized orbitals, such as the $3d$ or $4f$
states, and then treat the interaction Hamiltonian explicitly in a
proper way (beyond LDA). The total Hamiltonian reads:

\begin{eqnarray}
&&H^{LDA+G}=H^{LDA}+H_{int}-H_{dc}     \nonumber  \\
&&H_{int}=U\sum_{i\alpha} n_{i\alpha}^{\uparrow} n_{i\alpha}^{\downarrow}
+\frac{U^\prime}{2}\sum_{\stackrel{\alpha\neq\beta}{i,\sigma,\sigma^\prime}}
n_{i\alpha}^\sigma n_{i\beta}^{\sigma^\prime}-\frac{J}{2}
\sum_{i\sigma,\alpha\neq\beta} n_{i\alpha}^\sigma n_{i\beta}^\sigma \nonumber \\
&&<H_{dc}>^{LDA}=\overline{U}N(N-1)-\frac{\overline{J}}{2}
\sum_{\sigma}[N^\sigma(N^\sigma-1)]
\end{eqnarray}

\noindent where $\left\vert i\alpha \right\rangle^\sigma $ are a
set of local orbitals with spin index $\sigma$ and occupation
number $n_{i\alpha}$ for lattice site $i$; the $U, U^\prime$ and
$J$ gives the intra-, inter-orbital repulsive interaction and
Hund's exchange coupling, respectively. The $H_{dc}$ is the double
counting term from LDA, where interaction strength $\overline{U}$
and $\overline{J}$ are averaged over orbitals.

If the interaction term is treated by the Hartree-like scheme
(LDA+$U$), the correction over LDA is a set of energy shift of the
local orbitals, leaving the kinetic part unchanged. This is fine
if the fluctuation effect is not strong, but will fail in opposite
case, such as the Na$_x$CoO$_2$ system studied here. In this
sense, the DMFT method, in which frequency-dependent self-energy
is properly computed, is much better than LDA+$U$. However, due to
the heavy computational cost for multi-orbtial systems, the
current DMFT studies~\cite{Ishida,DMFT-new} are all applied to
tight-binding Hamiltonians extracted from LDA without full charge
density self-consistency. This is insufficient if the charge, spin
and orbtial degrees of freedom are all active as discussed above
for Na$_x$CoO$_2$.

In the LDA+Gutzwiller approach~\cite{LDA+G}, the Gutzwiller wave
function $|\Psi _{G}\rangle =\hat{P}|\Psi _{0}\rangle $ ($\hat{P}$
is a projection to many-body configuration) is used instead of
single slater determinate wave function $|\Psi _{0}\rangle $. The
orbital, charge and spin fluctuations can be included by the
multi-configuration nature of the Gutzwiller wave function. As the
results, a set of orbital-dependent kinetic energy renormalization
factor $Z_{\alpha }$ are obtained for the correlated states in
addition to the on-site energy shift. Unlike the previous
Gutzwiler or DMFT studies~\cite{ZQWang,Ishida,DMFT-new}, here all
charge transfer processes, crystal field and orbital fluctuations
are self-consistently treated within the framework of DFT, which
allows for the accurate computation of ground state total energy.

We use the plane-wave pseudo-potential method, and choose the
Co-$3d$ wannier functions as the correlated local orbitals. The
atomic-limit convention $U=U^\prime+2J$ is followed. How to
determine the value of $U$ and $J$ is a common problem for
LDA+$U$, DMFT and present methods, and no unified way is
established yet.  Nevertheless, reasonable estimations have been
done for $U=3.0\sim 5.0 eV$ and $J\sim 1.0 eV$ for Na$_x$CoO$_2$
system following the literatures~\cite{ZQWang,Ishida}. Instead of
using single fixed $U$, various values have been studied, and our
qualitative results are not changed, as shown below. In addition,
since our main purpose is to establish a general picture for the
physics of Na$_x$CoO$_2$, the structure differences among
different doping $x$ are neglected, and the Na doping is treated
by virtual crystal approximation (therefore the charge-ordered
states with Na ordering at particular doping, which is interesting
but not our purpose here, is out of the phase diagram).

\begin{figure}
\includegraphics[scale=0.43]{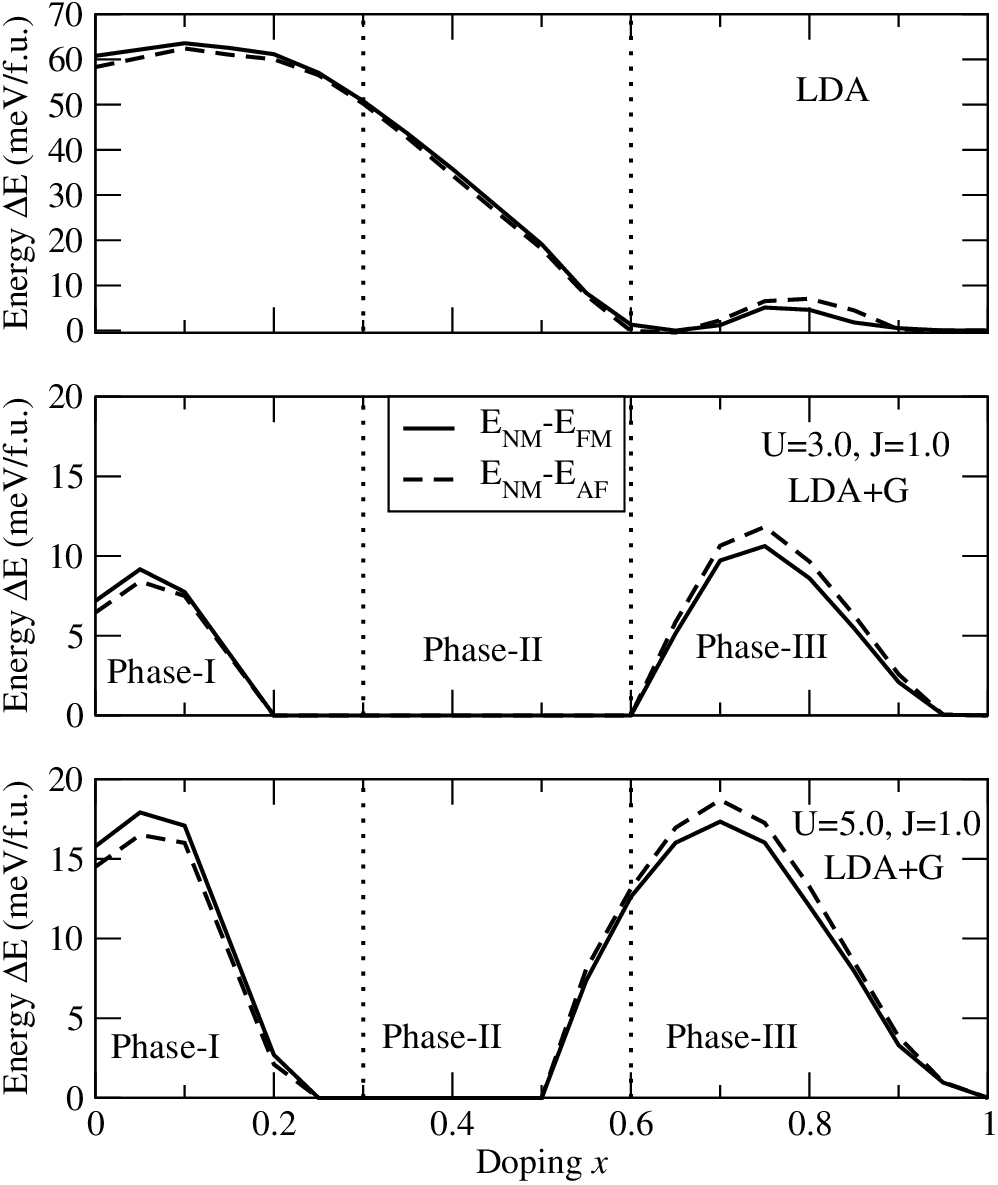}
\caption{The calculated stabilization energies of FM (solid lines)
and layer-type AF (dashed lines) states with respect to NM state
for Na$_x$CoO$_2$ over the whole doping range $0.0<x<1.0$. (a)
Results obtained by LDA; (b) LDA+Gutzwiller results obtained using
$U=3.0$ eV and $J=1.0$ eV; (c) LDA+Gutzwiller results obtained
using $U=5.0$ eV and $J=1.0$ eV.}
\end{figure}

Fig.1 shows the phase diagram computed for the whole doping range
0.0$<$$x$$<$1.0. The solid lines and the dashed lines represent
the stabilization energies of FM state and layer-type AF state
relative to NM solution, respectively. The intra-plane AF state is
hard to be stabilized due to geometrical fluctuation of triangle
lattice. The LDA (shown in Fig.1(a)) gives magnetic ground states
for all doping $x$, which are inconsistent with experiments. In
contrast, the phase diagram by LDA+Gutzwiller has three distinct
regions, which can be understood as the consequence of competition
among crystal field splitting, inter-orbtial charge and spin
fluctuation, as discussed in the following parts. As shown in
Fig.1(b) and (c), the features of phase diagram are qualitatively
the same using $U$=3.0 eV or $U$=5.0 eV.

{\noindent \it  Phase II: correlated non-magnetic metal (0.3$<$$x$$<$0.6)}.

First of all, the NM state is now correctly predicted for this
region. It is known that LDA overestimates the tendency to be FM
for several systems, such as ruthenates~\cite{Singh-Ru}, and this
artifact is even enhanced by LDA+$U$. The physical reason is that
correlation effect is not properly treated in LDA and LDA+$U$, but
it is well included in the present formalism. In Fig.2 we
summarize the properties of the NM solutions for the whole doping
range. For $x$ larger than $0.3$, as shown in Fig.2 (a), the
$e_g'$ bands are fully occupied and the inter-orbital fluctuation
(defined as F and S, see caption of Fig.2) is weak, indicating a
effective single band system. However with $x$ approaching the
phase boundary around $x_c=0.3$, a crossover to multi-band
behavior has been detected from the strong inter-orbital spin and
charge fluctuation as shown in Fig.2(b). Such fluctuations in low
doping area is induced by the Hund's rule coupling, which favors
even distribution of electrons among different orbitals with same
spin. Due to the presence of the correlation effect, the
quasi-particle band width (kinetic energy) is renormalized by
factor $Z_\alpha^2$ which is about 0.5 (0.7) for the $a_{1g}$
($e_g^\prime$) state at $x$=0.3 (as shown in Fig.2). The same
amplitude of renormalization is reported by ARPES~\cite{ARPES}.

\begin{figure}
\includegraphics[clip,scale=0.43]{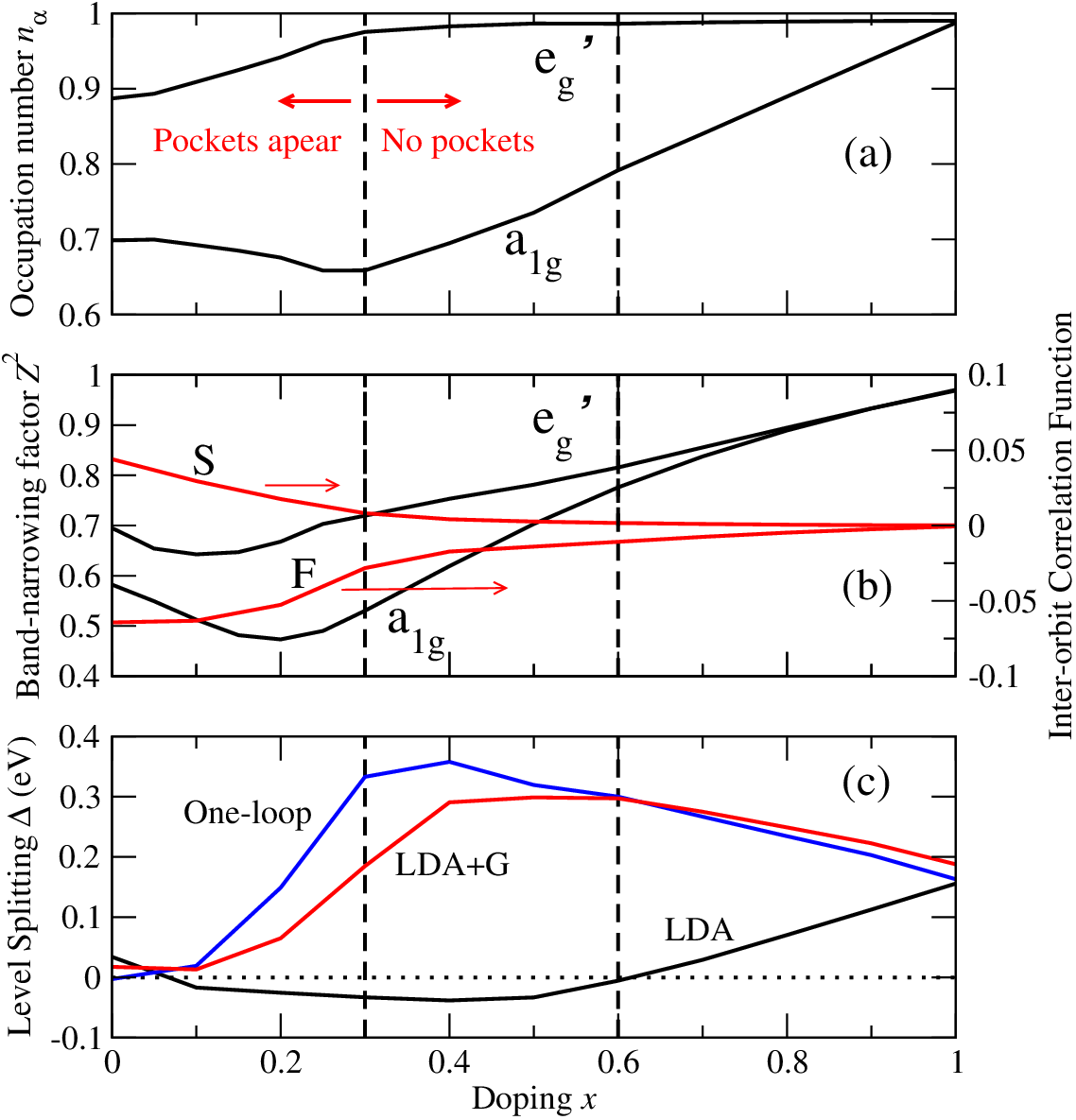}
\caption{Results for NM state as the functions of doping $x$ with
$U$=5.0eV and $J$=1.0eV. (a) Occupation numbers; (b) Band
renormalization $Z^2$-factors, inter-orbit charge fluctuation
F=$\langle\hat{n}_{a_{1g}}\hat{n}_{e_g^\prime}\rangle
-\langle\hat{n}_{a_{1g}}\rangle\langle\hat{n}_{e_g^\prime}\rangle$,
inter-orbit spin fluctuation
S=$\langle\hat{S}^z_{a_{1g}}\hat{S}^z_{e_g^\prime}\rangle
-\langle\hat{S}^z_{a_{1g}}\rangle\langle\hat{S}^z_{e_g^\prime}\rangle$;
and (c) Level splitting
$\Delta=\varepsilon_{a_{1g}}-\varepsilon_{e_g^\prime}$.}
\end{figure}

Secondly, it has long been a controversial issue whether the
$e_g^\prime$ states cross the Fermi level? To answer this question,
the correlation renormalized level shift is crucial (as shown in
Fig.2(c) the $a_{1g}$-$e_g^\prime$ level splitting $\Delta$ is much
renormalized compared to LDA results). Unfortunately after including
the correlation effect, two studies have been done, and conflicting
results are drawn~\cite{ZQWang,Ishida} for $x$=0.3. It was recently
pointed out by Marianetti {\it et al.} that the controversial is due
to the different choice of crystal-field splitting in their
tight-binding model~\cite{DMFT-new}. To go further, in our studies,
not only the crystal-field is treated parameter-free, but also the
full charge self-consistency is achieved. To see the difference, we
performed one-loop calculations (i.e. the charge-density is fixed to
the LDA value and only Gutzwiller wave functions are optimized), then
Marianetti's results are recovered, i.e, the $e_g^\prime$ pockets are
not present for $x$=0.3. In addition, we also found that even for
$x$=0.2 the $e_g^\prime$ pockets are not present in this one-loop
calculation. However, after including the charge density
self-consistency, the renormalization of level splitting are
suppressed (see Fig.2(c)). As the results, we found that $x$=0.3 is
the critical point, i.e. $e_g^\prime$ pockets are absent for $x>$0.3
but present for $x<$0.3.  Our calculated quasi-particle bands (as
shown in Fig.3 and discussed below) can be well compared with
ARPES. Since with the interactions we used ($U$=5.0eV and $J$=1.0eV)
the static Hartree Fock shift is zero~\cite{Ishida}, the renormalized
energy level shift here is fully contributed by the fluctuation
effect. From Fig.2(c), we found that the $\Delta$ is peaked in the
crossover region indicating that the inter-orbital fluctuation is the
main reason for the renormalization effect on the energy levels.

\begin{figure}
\includegraphics[clip,scale=0.35]{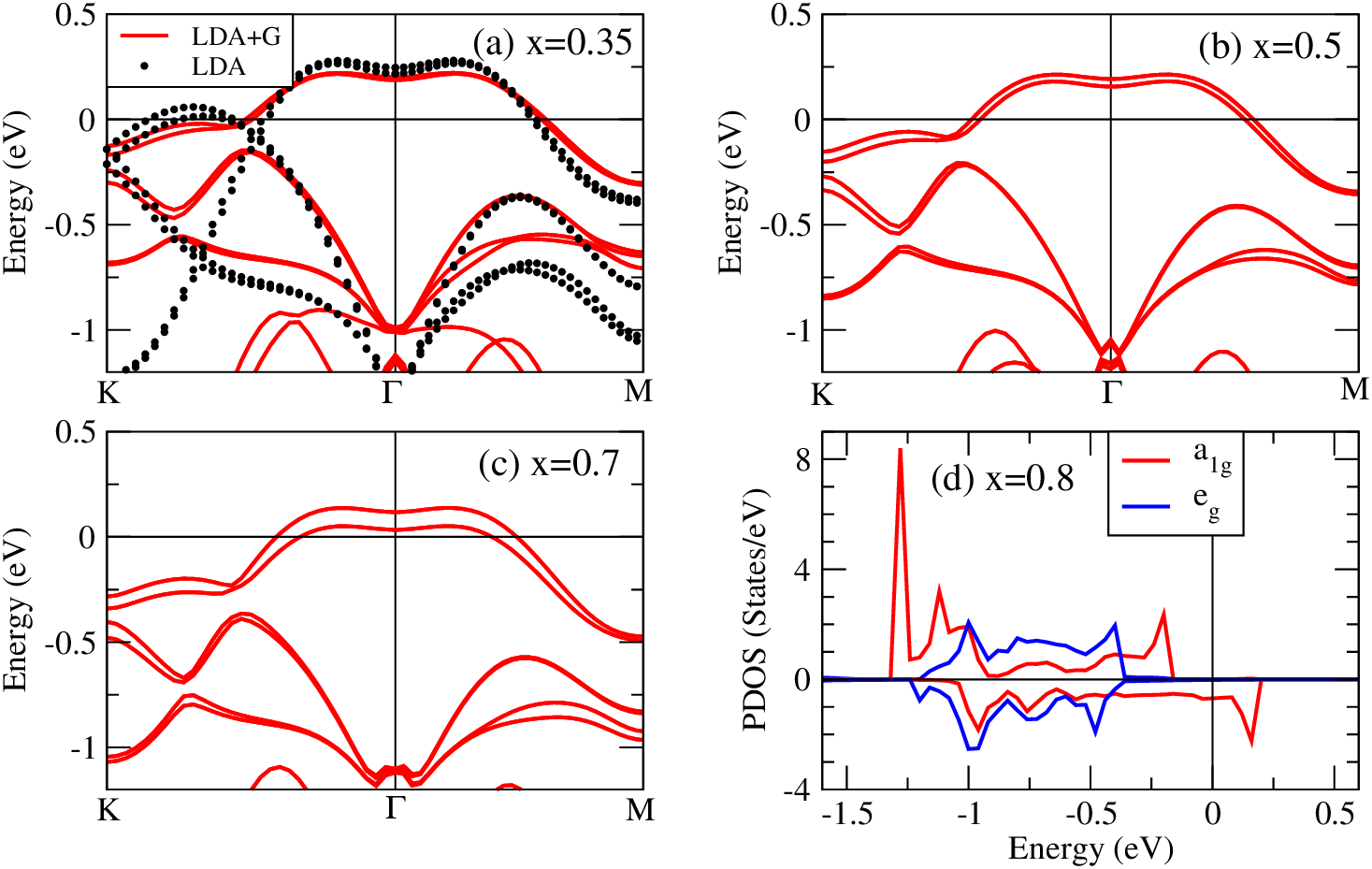}
\caption{The calculated band structure of Na$_x$CoO$_2$ for
(a)$x$=0.35; (b)$x$=0.5; (c)$x$=0.75 in the NM state. (d) is the
projected density of states (PDOS) for $x$=0.8 in the A-type AF
state. $U$=5.0 eV and $J$=1.0 eV are used, and the spin-orbital
coupling is included.}
\end{figure}

{\it Phase III: weakly correlated Stoner metal for $x>0.6$.}
Clearly seen from the calculated density of states (Fig.3(d)), a
sharp van-Hove singularity (VHS) is present near the $a_{1g}$ band
top-edge (due to the flat dispersion).  For the Na rich side, the
Fermi level is shifted close to the VHS, the Stoner instability
make the system FM (in-plane). This conclusion is supported by the
following facts for this region: (1) LDA works qualitatively well;
(2) the magnetic solution only weakly depends on interaction
strength $U$ and $J$ (see Fig.1); (3) calculated spin and charge
fluctuation are all weak (Fig.2(b)). It is therefore suggested
that strong correlation is not the driving force for the magnetic
state, instead the VHS is responsible. Furthermore, we correctly
predict that the A-type AF state is more stable than FM state (see
the difference between solid and dashed lines in Fig.1). From
Fig.3(d) for $x$=0.8, it is seen that spin moment mostly comes
from the $a_{1g}$ states, which aligns towards the $z$ direction
(inter-layer) of the crystal. According to the Goodenough-Kanamori
rule, the exchange-coupling along $z$ direction is dominates by AF
super-exchange, which stabilizes the A-type AF state. In fact, the
estimated inter-layer exchange coupling (from the total energy
difference between FM and A-AF solutions) is about $J_c$=3.0 meV,
in good agreement with experimental results~\cite{A-AF}. From
itinerant FM theory, the size of spin moment may change with
raising temperature, but in the presence of strong VHS near band
edge, this possibility is prevented by the sharp density barrier.
Indeed we found that the calculated moments of FM and A-type AF
solutions are the same. Therefore, the $a_{1g}$ moment will
behavior like localized spin as observed experimentally. In other
words, the Curie-Weiss behavior in this region does not
necessarily suggest the {\it enhanced correlation}.

{\it Phase I: magnetic correlated metal:} The main difference between
phase I and II is that the $e_g^\prime$ band start to go cross the
Fermi level and the $e_g^\prime$ hole pockets are present.  Therefore
phase I is effectively a multi-band system. In the meanwhile, the
magnetic instability is recovered, and the system is stabilized in FM
ground state. Several points should be addressed here to understand
this phase: (1) in contrast to phase III, the FM state is slightly
more stable than A-type AF state; (2) both $e_g^\prime$ and $a_{1g}$
contribute to the spin moment, in other words, the stabilization of FM
state is due to the enhancement of inter-orbit (rather than
intra-orbit) spin fluctuation (Fig.2(b)), which is induced by the
Hund's rule coupling; (3) strong correlation (rather than VHS) is
responsible for the magnetic instability. It is interesting to note
that the critical point $x\sim 0.3$ (boundary between phase I and II)
is close to the doping level where superconductivity was observed. The
experimental information for this region is quite limited and not
conclusive due to the difficulty of sample preparation, our prediction
should be evaluated by future experiments.

Finally, we show systematically in the Fig.3(a)-(c) the calculated
quasi-particle band dispersion for $x$=0.35, 0.5, and 0.7 after
including the spin-orbital-coupling (SOC) effect. The overall picture
can be nicely compared to ARPES data~\cite{ARPES}. In particular, (1)
the band width renormalization around factor of 2 is now obtained; (2)
the $e_g^\prime$ Fermi surface pocket is absent; (3) the
$a_{1g}$-$e_g^\prime$ anti-crossing along $\Gamma-K$ line is nothing
but a effect of SOC, and the gap around 0.1 eV is comparable to
experimental data~\cite{Anti}.

In summary, using the recently developed LDA+Gutzwiller method, we
are now able to calculate the ground state total energy of
correlated multi-orbital systems from {\it ab-initio} after taking
into account the orbital fluctuation. The calculated phase diagram
of Na$_x$CoO$_2$ establishes a general understanding for the
physics behind. Most of the discrepancies between experiments and
previous theories, such as the $e_g^\prime$ pocket,
$a_{1g}$-$e_g^\prime$ anti-crossing, Curie-weiss behavior for
$x>$0.5, are self-consistently understood. Three distinct phase
regions are identified, which is instructive to future
experiments.

We acknowledge valuable discussions with H. Ding, and the supports
from NSF of China (No.10334090, 10425418, 60576058), and that from the
973 program of China (No.2007CB925000).


\begin{references}


\bibitem{NaCoO_phase} M. L. Foo and,{\it et al.}, Phys. Rev. Lett
{\bf 92}, 247001 (2004).

\bibitem{Spin} Y. Wang, N. S. Rogado, R. J. Cava, N. P. Ong, Nature,
{\bf 423}, 425 (2003).

\bibitem{A-AF} J. Sugiyama et al., Phys. Rev. B 67, 214420 (2003);
A.T. Boothroyd, et al., Phys. Rev. Lett. {\bf 92}, 197201 (2004);
S. P. Bayrakci et al., Phys. Rev. Lett. 94, 157205 (2005);
L. M. Helme, Phys. Rev. Lett. {bf 94}, 157206 (2005).

\bibitem{SC} K. Takada, H. Sakurai, E. T. Muromachi, F. Izumi,
R. A. Dilanian, and T. Sasaki, Nature, {\bf 422}, 53 (2003).

\bibitem{x=0.5} K.-W. Lee and W. E. Pickett, Phys. Rev. Lett. {\bf
96}, 096403 (2006); G. Gasparovic, and et.al, Phys. Rev. Lett. {\bf
96}, 046403 (2006).

\bibitem{Singh} D. J. Singh, Phys. Rev. B {\bf 61}, 13397 (2000).

\bibitem{ARPES} H. B. Yang et al., Phys. Rev. Lett. 95, 146401 (2005);
D. Qian et al., Phys. Rev. Lett. 96, 046407 (2006); D. Qian et al.,
Phys. Rev. Lett. 96, 216405 (2006); D. Qian, L. Wray, D. Hsieh,
L. Viciu, R. J. Cava, J. L. Luo, D. Wu, N. L. Wang, and M. Z. Hasan,
Phys. Rev. Lett. 97, 186405 (2006).

\bibitem{Louie} P. Zhang and {\it et al.} Phys. Rev. Lett {\bf
93},236402 (2004); P. Zhang and {\it et al}, Phys. Rev. B {\bf 70},
085108 (2004).

\bibitem{ARPES2} M.Z. Hasan and {\it et al.} Phys. Rev. Lett {\bf
92},246402 (2004); H.B.Yang and {\it et al.} Phys. Rev. Lett {\bf
92},246403 (2004)


\bibitem{ZQWang} S. Zhou and {\it et al.} Phys. Rev. Lett {\bf
94},206401 (2005).

\bibitem{Ishida} H. Ishida, M. D. Johannes, and A. Liebsch3,
Phys. Rev. Lett.{\bf 94}, 196401 (2005).

\bibitem{DMFT-new} C. A. Marianetti, and {\it et al.},
Phys. Rev. Lett. {\bf 99}, 246404 (2007).

\bibitem{LDA+G} X. Y. Deng, X. Dai, and Z. Fang, cond-mat/0707.4606.

\bibitem{Anti} J. Geck, et. al. Phys. Rev. Lett. {\bf 99}, 046403
(2007).

\bibitem{DMFT} G. Kotliar, et al. Rev. Mod. Phys. {\bf 78}, 865
(2006).

\bibitem{Singh-Ru} I. I. Mazin, and D. J. Singh, Phys. Rev. B {\bf
56}, 2556 (1997); Z. Fang, and Terakura, {\bf 64}, 020509 (2001).




\end{references}
\end{document}